\documentclass[aps,prl,showpacs,twocolumn,amsmath,10pt,superscriptaddress,floatfix,footinbib]{revtex4-1}
\usepackage{epsfig,amssymb,amsfonts,bm}
\usepackage{array}
\usepackage[active]{srcltx}
\usepackage{color}
\newcolumntype{L}{>{$}l<{$}} 
\newcolumntype{C}{>{$}c<{$}} 
\renewcommand{\arraystretch}{1.2}

\newcommand{\be}{\begin{equation}}
\newcommand{\ee}{\end{equation}}
\newcommand{\bea}{\begin{eqnarray}}
\newcommand{\eea}{\end{eqnarray}}
\newcommand{\beas}{\begin{eqnarray*}}
\newcommand{\eeas}{\end{eqnarray*}}

\newcommand{\keV}{\,\mbox{keV}}
\newcommand{\MeV}{\,\mbox{MeV}}
\newcommand{\GeV}{\,\mbox{GeV}}
\newcommand{\al}{&}

\newcommand{\A}{\mathcal{A}}
\newcommand{\BR}{\mathcal{B}}
\newcommand{\nno}{\nonumber\\}
\newcommand{\spp}{s^\prime}

\newcommand{\disc}{\text{disc}}
\newcommand{\ints}{\int_{s_\text{th}}^\infty}

\newcommand{\mathf}{\mathcal{F}}
\newcommand{\matha}{\mathcal{A}}

\newcommand{\itp}{\affiliation{CAS Key Laboratory of Theoretical Physics,
            Institute of Theoretical Physics,\\ Chinese Academy of Sciences,
            Beijing 100190, China}}
\newcommand{\bonn}{\affiliation{Helmholtz-Institut f\"ur Strahlen- und
             Kernphysik and Bethe Center for Theoretical Physics,\\
             Universit\"at Bonn,  D-53115 Bonn, Germany}}
\newcommand{\fzj}{\affiliation{Institute for
           Advanced Simulation, Institut f\"ur Kernphysik and
           J\"ulich Center for Hadron Physics,\\ 
           Forschungszentrum J\"ulich, D-52425 J\"ulich, Germany}}
\newcommand{\ucas}{\affiliation{School of Physical Sciences,
            University of Chinese Academy of Sciences,
            Beijing 100049, China}} 
\newcommand{\tsu}{\affiliation{Tbilisi State University, 0186 Tbilisi, Georgia}}

\begin{document}
\title{
Where is the lightest  charmed scalar meson?
}

\author{Meng-Lin~Du} \email{du@hiskp.uni-bonn.de}\bonn
\author{Feng-Kun~Guo}\email{fkguo@itp.ac.cn}\itp\ucas
\author{Christoph~Hanhart} \email{c.hanhart@fz-juelich.de}\fzj
\author{Bastian~Kubis} \email{kubis@hiskp.uni-bonn.de}\bonn
\author{Ulf-G.~Mei{\ss}ner} \email{meissner@hiskp.uni-bonn.de}\bonn\fzj\tsu

\begin{abstract}

The lightest charmed scalar meson is known as the $D_0^*(2300)$, which is one of the earliest new hadron
resonances observed at modern $B$ factories. 
We show here that the parameters assigned to the lightest scalar $D$-meson are in conflict with 
the precise LHCb data of the decay $B^-\to D^+ \pi^- \pi^-$. On the contrary, these data
 can be well described by an unitarized chiral amplitude containing a much lighter charmed scalar meson,
the  $D_0^*(2100)$. 
We also extract the low-energy $S$-wave $D\pi$ phase of the decay $B^-\to D^+ \pi^- \pi^-$ from the data
in a model-independent way, and show that its difference from the $D\pi$ scattering phase shift can be traced back to
an intermediate $\rho^-$ exchange.
Our work highlights that an analysis of data consistent with chiral symmetry, unitarity, and analyticity
is mandatory in order to extract the properties of the ground-state scalar
mesons in the singly heavy sector correctly, in analogy to the light scalar mesons $f_0(500)$ and $K_0^*(700)$.

\end{abstract}

\maketitle

{\it Introduction.}---{\spaceskip=0.22em\relax Since the discovery of the $D_{s0}^*(2317)$~\cite{Aubert:2003fg}}, 
many hadrons were observed beyond the quark model expectations,
which have seriously challenged the understanding of the hadron spectrum in terms
of the conventional quark model that identifies mesons as $\bar qq$ states. The
observation that the $D_{s0}^*(2317)$~\cite{Aubert:2003fg} and $D_{s1}(2460)$~\cite{Besson:2003cp} are significantly
lighter than expected by the quark model, around $2.48$ and $2.55\GeV$~\cite{Godfrey:1985xj,Godfrey:2015dva,Ebert:2009ua}, 
has driven the development of various models, including $D^{(\ast)}K$ hadronic molecules~\cite{Barnes:2003dj,vanBeveren:2003kd,Szczepaniak:2003vy,Kolomeitsev:2003ac,Chen:2004dy,Guo:2006fu,Guo:2006rp,Gamermann:2006nm}, tetraquark states~\cite{Cheng:2003kg,Maiani:2004vq}, and mixtures of $c\bar{q}$ with tetraquarks~\cite{Browder:2003fk}. 
In 2004, two new charm-nonstrange structures, the $D_0^*(2300)$~\cite{Abe:2003zm,Link:2003bd}, called $D_0^*(2400)$
previously, and $D_1(2430)$~\cite{Abe:2003zm}, were reported as the SU(3) partners of the $D_{s0}^*(2317)$
and $D_{s1}(2460)$, respectively. The observations posed a puzzle: why are the masses of the
two nonstrange mesons, $D_0^*(2300)$ and $D_1(2430)$, almost equal to their strange siblings, i.e.,
the $D_{s0}^*(2317)$ and $D_{s1}(2460)$?
Thanks to new data from both lattice quantum chromodynamics (QCD)~\cite{Liu:2012zya,Mohler:2013rwa,Lang:2014yfa,Moir:2016srx,Bali:2017pdv,Cheung:2020mql} and the LHCb experiment~\cite{Aaij:2016fma}, it
was recently demonstrated that the various puzzles in the charm meson spectrum can be solved naturally in
the framework of unitarized chiral perturbation theory (UChPT)~\cite{Albaladejo:2016lbb,Du:2017zvv,Guo:2017jvc}
that allows one to calculate
 the nonperturbative dynamics of Goldstone bosons scattering off the $D_{(s)}^{(*)}$ mesons in a controlled way. 
The combination of UChPT and lattice QCD not only reproduced the correct $D_{s0}^*(2317)$ mass~\cite{Liu:2012zya},
but also predicted its pion mass dependence~\cite{Du:2017ttu,Bali:2017pdv}. 
The solution provided for the SU(3) mass hierarchy puzzle mentioned above is that
instead of only one heavy state, $D_0^*(2300)$ in the channel
$(S,I)\equiv(\text{strangeness}, \text{isospin})=(0,1/2)$,
there are two states, one lighter and one heavier~\cite{Kolomeitsev:2003ac,Guo:2006fu,Gamermann:2006nm,Guo:2009ct,Liu:2012zya,Guo:2015dha,Albaladejo:2016lbb,Guo:2018tjx,Guoa:2018dhm,Guo:2018kno}. 
The most recent studies revealed their pole locations to be at
$\left(2105^{+6}_{-8}-i\,102^{+10}_{-11}\right)$ and
 $\left(2451^{+35}_{-26}-i\, 134^{+7}_{-8}\right)\text{MeV}$~\cite{Albaladejo:2016lbb,Du:2017zvv}, respectively.
The SU(3) partner of the $D_{s0}^*(2317)$ is the lighter one, denoted as $D_0^*(2100)$ in the
following, which restores the expected mass hierarchy. 
The heavier pole on the other hand is a member of a different multiplet.
Support for the presence of two poles comes from
an analysis of the high-quality
LHCb data on the decays $B^-\to D^+\pi^-\pi^-$~\cite{Aaij:2016fma}, $B_s^0\to\bar{D}^0K^-\pi^+$~\cite{Aaij:2014baa},
$B^0\to \bar{D}^0 \pi^-\pi^+$~\cite{Aaij:2015sqa}, $B^-\to D^+\pi^- K^-$~\cite{Aaij:2015vea}, and
$B^0\to\bar{D}^0\pi^- K^+$~\cite{Aaij:2015kqa} performed
in Refs.~\cite{Du:2017zvv,Du:2019oki}, as well as from the fact that their existence is consistent
with the lattice energy levels~\cite{Liu:2012zya,Mohler:2013rwa,Lang:2014yfa,Moir:2016srx,Bali:2017pdv} for the relevant two-body scattering~\cite{Albaladejo:2016lbb,Albaladejo:2018mhb,Guo:2018tjx}.
This two-pole structure indeed emerges as a more general pattern in the hadron spectrum, see, e.g.,
Ref.~\cite{Meissner:2020khl}.

Despite the  phenomenological success of this picture in describing the available lattice and LHCb data, the observation
that  the lightest $D_0^*$ has a mass around $2.1\GeV$ has not entered the Review of Particle Physics
(RPP)~\cite{Zyla:2020zbs} yet, which still lists the $D_0^*(2300)$ as the lightest charmed scalar meson and the $D_1(2430)$
as the corresponding axial-vector meson.
In this Letter, we demonstrate that the $D_0^*(2300)$ as in the RPP is not consistent with the most precise data
for $B^-\to D^+\pi^-\pi^-$, contrary to the $D_0^*(2100)$ predicted in UChPT, and conclude that the
positive-parity charm-nonstrange meson spectrum in the RPP needs to be revised.

{\it $D\pi$ $S$-wave phase of $B^-\to D^+\pi^-\pi^-$.}---The decay amplitude in the low-energy region of the
$D\pi$ system can be decomposed into $S$-, $P$-, and $D$-waves,
\bea\label{eq:pwdecomposition}
\matha_{B^-\to D^+\pi^-\pi^-}(s,z) = \sum_{\ell=0}^2 \sqrt{2\ell+1} \matha_\ell(s) P_\ell (z_s) \,,
\eea
where $\matha_\ell(s)$ with $\ell = 0,1,2$ correspond to the amplitudes with $D^+\pi^-$ in the $S$-, $P$-, and
$D$-waves, respectively, $s$ is the c.m.\ energy squared of the $D^+\pi^-$ system, and $P_\ell(z_s)$ are the
Legendre polynomials with $z_s$ the cosine of the helicity angle of the $D^+\pi^-$ system, i.e.,
the angle between the moving directions of the two pions in the $D^+\pi^-$ c.m.\ frame. The angular
moments are determined by weighting the data with the Legendre polynomials $P_\ell(z)$~\cite{Aaij:2016fma}.
They contain contributions from certain partial waves and their interference terms, and thus the
corresponding phase variations. The first few moments are given by~\cite{Aaij:2016fma,Du:2017zvv,Du:2019oki}
\begin{align}
  \langle P_0\rangle &\propto |\A_0|^2 + |\A_1|^2 + |\A_2|^2 \,,\nonumber\\
  \langle P_2\rangle &\propto \frac25|\A_1|^2 + \frac27|\A_2|^2 
+\frac{2}{\sqrt{5} } |\A_0||\A_2| \cos(\delta_2-\delta_0) \, ,\nonumber\\
\langle P_{13}\rangle &\equiv \langle P_1\rangle -\frac{14}{9} \langle 
P_3\rangle \propto \frac2{\sqrt{3}} |\A_0||\A_1| \cos(\delta_1-\delta_0) \,,
\label{eq:moments}
\end{align}
with $\delta_i$ the phase of $\matha_i$, i.e., $\matha_i = |\matha_i|e^{i\delta_i}$. As first proposed in
Ref.~\cite{Du:2017zvv}, we use the linear combination $\langle P_{13}\rangle$ instead of $\langle P_1\rangle$
and $\langle P_3\rangle$ individually, since it only depends on the $S$-$P$-wave interference up to $\ell = 2$ and
is particularly sensitive to the $S$-wave phase motion. 

\begin{figure}[tb]
\begin{center}
\includegraphics[width=\linewidth]{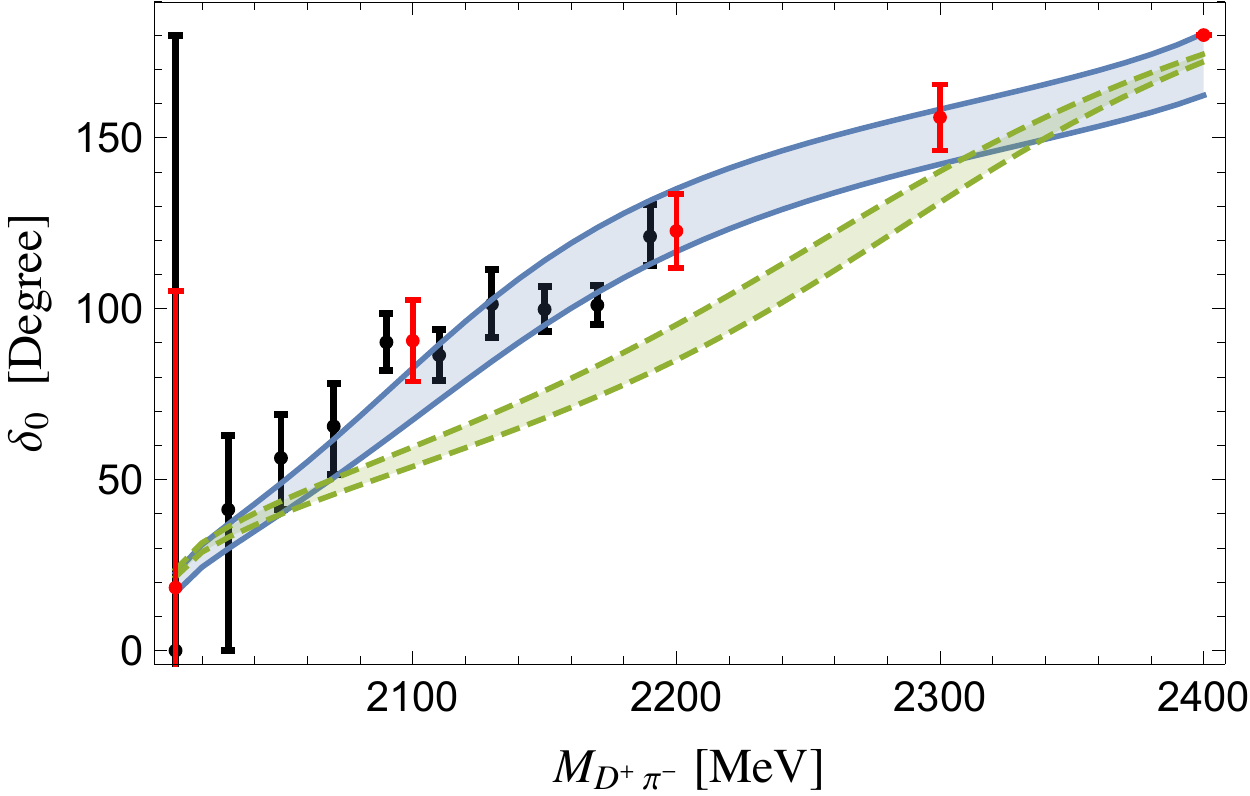}
\end{center}
\caption{
Comparison of the predictions of Eq.~\eqref{eq:A0pert} from UChPT (blue) and a Breit-Wigner parametrization (green) for $\delta_0$ with the phase extracted in
Ref.~\cite{Aaij:2016fma} (red) and that using Eq.~\eqref{eq:cos} (black). 
The bands  correspond to errors  propagated from the input UChPT scattering
amplitudes and from the Breit-Wigner resonance parameters.
}
\label{fig:pert}
\end{figure}

For $M_{D^+\pi^-} < 2.2\GeV$,  even the $D$-wave can be neglected, since the narrow
tensor resonance $D_{s2}(2460)$ is sufficiently far away. This can be verified from the data of the angular
moments, i.e., Fig.~3 in Ref.~\cite{Aaij:2016fma}. 
Therefore, in this kinematic regime one obtains 
\bea\label{eq:cos}
\cos (\delta_0-\delta_1) = \sqrt{\frac{3}{10}} \frac{\langle P_{13}\rangle}{\sqrt{
\langle P_2\rangle}\sqrt{\langle P_0\rangle - \frac{5}{2}\langle P_2\rangle}}
\eea
for the $S$-$P$ phase difference.
The $P$-wave is dominated by the vector resonance $D^*(2007)^0$ below the $D^+\pi^-$ threshold with a
width of less than $60\keV$~\cite{Rosner:2013sha,Guo:2019qcn}. The next vector $D^*$ resonance is far above this energy region. Thus, the phase of the $P$-wave $\delta_1$
can be safely fixed to $180^\circ$ for the region we are interested in. 
The $S$-wave $D\pi$ phase motion
below 2.2\GeV{} can then be extracted, 
see Fig.~\ref{fig:pert}. 
For comparison, the phase motion of the $D^+\pi^-$ $S$-wave up to 2.4\GeV{} obtained in the LHCb
analysis~\cite{Aaij:2016fma} (with the phase at 2.4\GeV{} fixed to $180^\circ$) is also shown, which is
fully in line with the phase we extracted from Eq.~\eqref{eq:cos} below 2.2\GeV. 

\begin{figure}[tb]
\begin{center}
\includegraphics[width=0.65\linewidth]{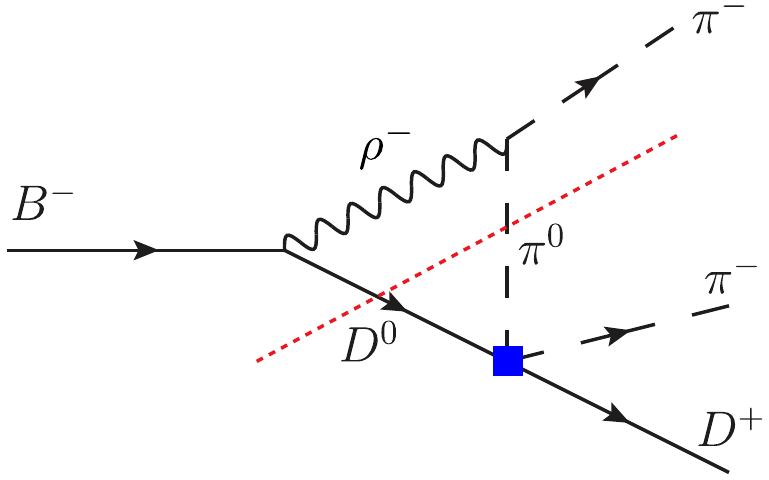}
\end{center}
\caption{The decay $B^-\to D^+\pi^-\pi^-$ via the coupled channel $B^-\to D^0\pi^0\pi^-$.
The filled square denotes the $D^0\pi^0\to D^+\pi^-$ $T$-matrix element.}
\label{fig:feyndiag}
\end{figure}

For $M_{D^+\pi^-}<2.4\GeV$, the effect of the $\rho$ meson could be significant
via the coupled channel $B^-\to D^0\pi^0\pi^-$; see Fig.~\ref{fig:feyndiag}. This follows directly from
the large branching ratio $\BR(B^-\to D^0\rho^-)=1.34\%$, which is an order larger than 
$\BR(B^-\to D^+\pi^-\pi^-)=0.107\%$~\cite{Zyla:2020zbs}. It is therefore reasonable to assume that the decay $B^-\to
D^+\pi^-\pi^-$ is dominated by the process $B^-\to D^0\rho^- \to D^0\pi^0\pi^-\to D^+\pi^-\pi^-$. By virtue
of soft-pion theorems, one has~\cite{Lin:1984ut}
\begin{align}
\A(B^-\to D^+\pi^-\pi^-)\big|_{p_{\pi^-}\to 0} &= \frac{1}{F_\pi} \A (B^0 \to \bar{D}^0\pi^0) \, ,\nno
\A(B^-\to D^0\pi^0\pi^-)\big|_{p_{\pi^0}\to 0} &= -\frac{1}{F_\pi} \A(B^-\to D^0\pi^-)\, ,
\end{align}
where $p_{\pi^-(\pi^0)}$ is the momentum of the $\pi^-(\pi^0)$, and $F_\pi$ is the pion decay constant (in the chiral limit). 
From $\BR(B^0\to \bar{D}^0\pi^0)=2.63\times 10^{-4}$ and $\BR(B^-\to D^0\pi^-)=4.68\times 10^{-3}$,
one concludes that
at low energies for the $D^+\pi^-(D^0\pi^0)$
system, the amplitude of $B^-\to D^0\pi^0\pi^-$ is much larger than that of $B^-\to D^+\pi^-\pi^-$.
Furthermore, isospin symmetry
shows that for the decays $B\to D\pi\pi$ with even relative angular momenta between
the pions, the amplitude for $B^-\to D^+\pi^-\pi^-$ is larger than that of $B^-\to D^0\pi^0\pi^-$ by a factor of $2\sqrt{2}$~\cite{Savage:1989ub,Niecknig:2015ija,Du:2017zvv,Du:2019oki}. 
As in addition even angular momenta here imply isospin $I=2$ and therefore nonresonant partial waves,
the relative angular momentum of $\pi^0\pi^-$ in the decay $B^-\to D^0\pi^0\pi^-$ is by far dominantly odd in the low-energy regime for $D^0\pi^0$, and the $\rho^-$
plays a crucial role. 

If we assume that the decay $B^-\to D^+\pi^-\pi^-$ is dominated by the process in Fig.~\ref{fig:feyndiag},
the $D\pi$ $S$-wave part of the triangle diagram can be estimated by the integral 
\bea\label{eq:A0pert}
\A_0^{\text{trig}} (s) = \frac{1}{\pi}\ints d\spp \frac{\hat P(\spp)\rho(\spp)
T_{D^0\pi^0\to D^+\pi^-}(\spp)}{\spp-s},
\eea
where $\hat{P}(s)$ is the production amplitude for $B^-\to D^0\rho^-\to D^0\pi^0\pi^-$ projected to
the $D^0\pi^0$ $s$-channel, $\rho(s)=\sqrt{\lambda(s,M_D^2,M_\pi^2)}/{(16\pi s)}$ is the $D\pi$ phase
space with $\lambda(a,b,c)=a^2+b^2+c^2-2ab-2ac-2bc$ the K\"all\'en function, $T_{D^0\pi^0\to D^+\pi^-}(s)$
the $S$-wave scattering amplitude for $D^0\pi^0\to D^+\pi^-$, and $s_{\text{th}}=(M_D+M_\pi)^2$. 
The expression for $\hat{P}(s)$ is the same as $\hat{\mathf}_0^{1/2}(s)$ in Eq.~\eqref{eq:hatF} below. 

The evaluation of Eq.~\eqref{eq:A0pert} depends on the asymptotic behavior of the integrand, which is
divergent in general. We may estimate Eq.~\eqref{eq:A0pert} using a cutoff 
at $\sqrt{s_\text{max}} = \sqrt{q_\text{max}^2+M_D^2}+\sqrt{q_\text{max}^2+M_\pi^2}$, where
$q_\text{max}\approx 1\GeV$ (another way is to introduce a form factor, e.g., $e^{-(s-s_\text{th})/s_0}$ with
$s_0 = \mathcal{O}(1\GeV)$~\cite{Szczepaniak:2015eza}). 
We evaluate Eq.~\eqref{eq:A0pert} by employing both the $D\pi$ scattering amplitude from UChPT~\cite{Liu:2012zya} and that of a Breit-Wigner (BW) parametrization of the $D_0^*(2300)$ for comparison,
despite the deficiencies of the latter discussed in Ref.~\cite{Du:2019oki}; see also Ref.~\cite{Gardner:2001gc}.

The results with $q_\text{max}=1\GeV$ are shown in Fig.~\ref{fig:pert}, where the solid blue band and the
green dashed band correspond to the $D\pi$ scattering amplitudes from UChPT and BW, respectively. The
obtained phase describes the data perfectly for the UChPT amplitude, while the BW one fails. We have
checked that the obtained phases are insensitive to a variation of the cutoff in a reasonable region, $q_\text{max} \in [0.8, 1.2]\GeV$.

{\it Khuri-Treiman formalism.}---While Eq.~\eqref{eq:A0pert} provides a reasonable estimation of the $S$-wave
decay amplitude with a clear underlying physical picture, it does not respect three-body unitarity.
In order to check if the conclusion formulated above is robust, we cure this deficiency by
employing the Khuri-Treiman equations~\cite{Khuri:1960zz}, which are based on two-body elastic phase shifts and explicitly
generate the crossed-channel rescattering between final-state particles. The formulas are constructed from dispersion relations for the related crossed scattering processes and then analytically continued to
the decay region, referring to the continuation of the triangle graph~\cite{Bronzan:1963mby}.

We can write amplitudes for $\A_{+--}(B^-\to D^+\pi^-\pi^-)$ and
$\A_{00-}(B^-\to D^0\pi^0\pi^-)$ in terms of single-variable functions according to a reconstruction theorem~\cite{Niecknig:2015ija,Niecknig:2017ylb},
\bea
 \A_{+--}&(s,&t,u)  =  \mathcal{F}_0^{1/2}(s) + \frac{\kappa(s)}{4}z_s \mathcal{F}_1^{1/2}(s) \nno 
& + & \frac{\kappa(s)^2}{16}(3z_s^2-1)\mathcal{F}_2^{1/2}(s) + (t \leftrightarrow s)\, , \nno
\A_{00-}&(s,&t,u)  =  -\frac{1}{\sqrt{2}}\mathcal{F}_0^{1/2}(s)- \frac{\kappa(s)}{4\sqrt{2}}z_s \mathcal{F}_1^{1/2}(s) \label{eq:construction} \\
& - & \frac{\kappa(s)^2}{16\sqrt{2}}(3z_s^2-1)\mathcal{F}_2^{1/2}(s) +\frac{\kappa_u(u)}{4}z_u\mathcal{F}_1^1(u)\, , \nonumber
\eea
where the subindex $\ell$ and superindex $I$ of the single-variable amplitudes $\mathcal{F}_\ell^I$ represent
the angular momentum and isospin, respectively, and only the $I < 3/2$ 
and $\ell \leq 2$ terms are taken into
account. 
The Mandelstam variables of the $B$-meson decay $B^-(p_B)\to D(p_D)\pi(p_1)\pi^-(p_2)$ are
 $s=(p_B-p_2)^2$, $t = (p_B-p_1)^2$, and $u=(p_B-p_D)^2$. The corresponding angles are given by 
\be
z_s \equiv \cos\theta_s = \frac{s(t-u)-\Delta}{\kappa(s)} \,, \quad z_u \equiv \cos\theta_u = \frac{t-s}{\kappa_u(u)} \,,
\ee
where $\kappa(s)=\lambda^{1/2}(s,M_D^2,M_\pi^2)\lambda^{1/2}(s,M_B^2,M_\pi^2)$, $\kappa_u(u)
=\lambda^{1/2}(u,M_B^2,M_D^2)\sqrt{1-{4M_\pi^2}/{u}}$, and $\Delta=(M_B^2-M_\pi^2)(M_D^2-M_\pi^2)$.

Since we are interested in the $s$-channel process, we use the index $A$ ($B$) to label the two-body channels corresponding to $D^+\pi^-$ and $D^0\pi^0$.
The partial-wave decomposition for 
the decay amplitudes $\A_A$ reads
\bea
\A_A(s,z_s) = \sum_{I,\ell} b_{I,\ell}^A P_\ell(z_s) f_\ell^I(s) \,,
\eea
with 
$b_{I,\ell}^A$ denoting Clebsch-Gordan coefficients. By comparing with Eq.~\eqref{eq:pwdecomposition},
it is easy to obtain $\A_\ell(s) = (2\ell+1)^{-1/2} \sum_I b_{I,\ell}^1 f_\ell^I(s)$.
We have the following partial-wave unitarity relation for elastic rescattering:
\bea
\disc f_\ell^I(s) = 2i f_\ell^I(s) \sin \delta_\ell^I(s) e^{-i\delta_\ell^I(s)} \theta(s-s_\text{th}) \,,
\eea
where $\delta_\ell^I(s)$ is the elastic final-state scattering phase shift. The discontinuities of $f_\ell^I$ and
those of the single-variable functions $\kappa^\ell \mathf_\ell^I$ coincide on the right-hand cut 
by construction. Thus, one has 
\begin{equation}
  \text{disc} \mathf_\ell^I(s) = 2i\big[\mathf_\ell^I(s)+\hat{\mathf}_\ell^I(s)\big]\sin\delta_\ell^I(s)
  e^{-i\delta_\ell^I(s)}\theta(s-s_\text{th}) \,,
\end{equation}
where the inhomogeneities $\hat{\mathf}_\ell^I(s)$ encode the left-hand cut contributions and are free of
discontinuities on the right-hand cut. 
This discontinuity relation is solved by
\begin{equation}
  \label{eq:Finhomo}
  \mathf_\ell^I(s)=\Omega_\ell^I(s)\bigg\{ Q_\ell^I(s) +\frac{s^n}{\pi}\ints \frac{d\spp}{s^{\prime n}}
  \frac{\sin\delta_\ell^I(\spp)\hat{\mathf}_\ell^I(\spp)}{|\Omega_\ell^I(\spp)|(\spp-s)}\bigg\}, 
\end{equation}
where 
$\Omega_\ell^I(s)=\exp\big\{ {s}/{\pi} \int_{s_\text{th}}^\infty ds'{\delta_\ell^I(s') }/[s'(s'-s)] \big\}$ 
is the Omn\`es function~\cite{Omnes:1958hv}, $Q_\ell^I(s)$ is a polynomial at least of degree $(n-1)$
(see discussion below),
and the number of subtractions $n$ is chosen to guarantee the convergence of the dispersion integral.

The inhomogeneity $\hat{\mathf}_\ell^I$ is determined by the partial-wave decomposition of
Eq.~\eqref{eq:construction} as the projection of the crossed-channel amplitudes onto the considered channel.
Around the $D\pi$ threshold in the $s$-channel, $\sqrt{t}\sim 5\GeV$, there is no resonance in the $t$-channel,
and thus the interaction is supposed to be very weak. The only possible significant crossed-channel effect
is from the $\rho$ meson through $B^-\to D^0\pi^0\pi^-$. The resulting inhomogeneity for the $S$-wave 
$s$-channel amplitude is~\cite{Niecknig:2015ija}
\bea\label{eq:hatF}
\hat{\mathcal{F}}_0^{1/2}(s) = -\frac{1}{4\sqrt{2}}
\int_{-1}^1 dz_s (t-s) \mathcal{F}_1^1(u) \,.
\eea
For technical details regarding this integral, see Ref.~\cite{Kacser:1963zz} and the Supplemental Material~\cite{supp}.

The full solution for the decay amplitudes can be obtained by solving a set of coupled integral equations in terms of a few
linearly independent complex subtraction constants contained in $Q_\ell^I(s)$, which cannot be determined \textit{a priori} in the framework of
dispersion theory.
Since we are only interested in
the $D\pi$ low-energy regime and especially in its $S$-wave, 
based on the large branching ratio of $B^-\to D^0\rho^-$, it is reasonable to approximate
$\mathf_1^1(u)$ in Eq.~\eqref{eq:construction} by a BW function for the $\rho$ meson.
In this case, $\mathf_1^1(u)$ behaves as $u^{-1}$ for $u\to \infty$, thus $\hat{\mathf}_\ell^I(s)$ in
Eq.~\eqref{eq:hatF} approaches a constant as $s\to \infty$. The number of the subtractions $n$ in
Eq.~\eqref{eq:Finhomo} is then determined by the asymptotic behavior of the scattering phase $\delta_\ell^I(s)$.
For the BW phase and that of UChPT taken from Ref.~\cite{Liu:2012zya}, one single subtraction is sufficient.
Moreover, the phase of UChPT, as well as that of the BW, is unreliable at high energies. 
Thus, the dispersion integral will be evaluated up to a cutoff $\Lambda$, and the effect of cutting off
the integral may be absorbed into the polynomial $Q_\ell^I(s)$.
Explicitly, for an integral
\begin{align}
g(s) &= \ints d\spp \frac{f(\spp)}{\spp-s} = \int_{s_\text{th}}^\Lambda d\spp \frac{f(\spp)}{\spp-s}
+ \int_\Lambda^\infty d\spp \frac{f(\spp)}{\spp-s} \nno
& \approx  g_0+g_1s+ \int_{s_\text{th}}^\Lambda d\spp \frac{f(\spp)}{\spp-s} \,.
\end{align}
For simplicity, we neglect the $I=3/2$ contribution since it contains no resonances. Therefore, for the
$S$-wave amplitude at low energies, Eq.~\eqref{eq:Finhomo} can be written as
\begin{align}\label{eq:Fspcut}
\mathf^{1/2}_0(s)\al =  \Omega_0^{1/2}(s)\bigg\{ g_0+g_1\frac{s-M_D^2}{M_D^2} \nno 
\al \quad + \frac{s}{\pi}\int_{s_\text{th}}^\Lambda \frac{d\spp}{\spp} \frac{\sin \delta_0^{1/2}(\spp)
  \hat{\mathcal{F}}_0^{1/2}(\spp)}{|\Omega_0^{1/2}(\spp)|(\spp-s)} \bigg\} \,.
\end{align}
The constants $g_0$ and $g_1$ have to be fixed by data. 

We fit 
$\langle P_0\rangle$, $\langle P_{13}\rangle$, and $\langle P_2\rangle$
of the decay $B^-\to D^+\pi^-\pi^-$ up to 2.4\GeV, which is below the $D\eta$ and $D_s\bar K$ thresholds.
To describe the angular moments, one needs the explicit amplitudes for the $D\pi$ $P$- and $D$-waves.
The $P$-wave can be safely parametrized
as $\delta_1^{1/2}(s)=\pi\,\theta\big(s-M_{D^{*0}}^2\big)$ as discussed below Eq.~\eqref{eq:cos}.
Consequently, the
dispersion integral \eqref{eq:Finhomo} for the $P$-wave can be neglected since $\sin\pi=0$. The $D$-wave is
dominated by the resonance $D_2^0(2460)$ with a width of
$47.5\MeV$~\cite{Zyla:2020zbs}, which is above the region we are interested in. 
Thus, the $D$-wave phase is close to $0$ below 2.4\GeV, and the corresponding dispersion integral
can be neglected as well. 
Therefore, for the $P$-($D$-)wave amplitudes $\mathf_{1,2}^{1/2}$, we can use the same BW forms as those in the
LHCb analysis~\cite{Aaij:2016fma}, which is equivalent to the corresponding Omn\`es function multiplied by a
polynomial. 
In the isobar model used in Ref.~\cite{Aaij:2016fma}, complex factors are introduced for each resonance BW function. 
Without crossed-channel effects, these factors become real according to Watson's theorem~\cite{Watson:1954uc}.
For the $P$- and $D$-waves, as discussed above, the Omn\`es representation should be a good approximation in
the energy region we are interested in, and the normalization factor is real. We also consider a complex
normalization and find the results unchanged.

\begin{figure*}[tb]
\begin{center}
\includegraphics[width=\linewidth]{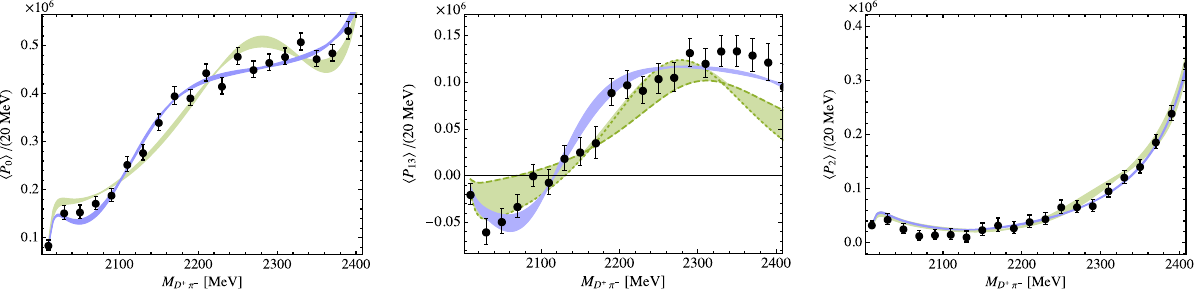}
\end{center}
\caption{Results of UChPT (blue) and the BW description (green), with the best fits $\chi^2/\text{d.o.f.}=1.2$ and 2.0, respectively.}\label{fig:KT}
\end{figure*}

\begin{figure}[tb]
\begin{center}
\includegraphics[width=\linewidth]{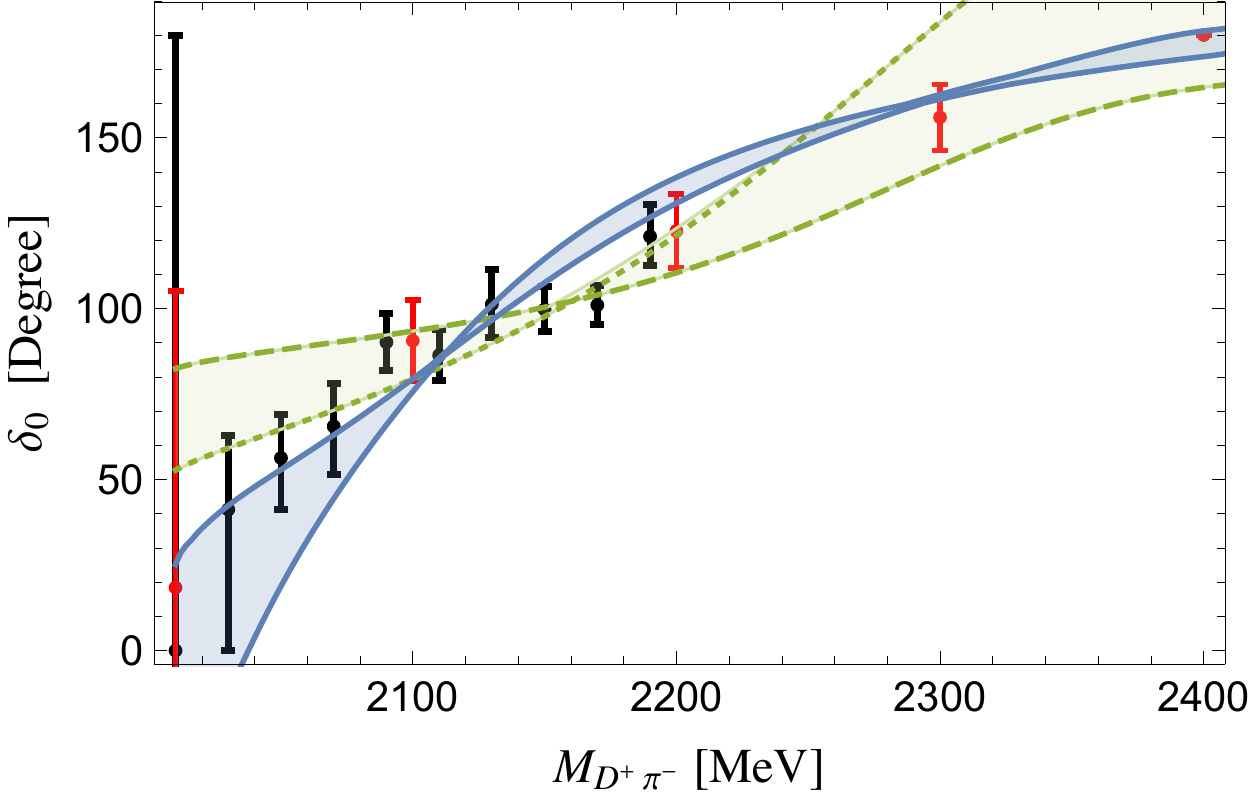}
\end{center}
\caption{Phases obtained from Eq.~\eqref{eq:Fspcut} with scattering phase shifts from UChPT (blue) and BW (green). 
  }
\label{fig:phaseIV}
\vspace{-0.2cm}
\end{figure}

For the $D\pi$ $S$-wave, we employ both the scattering phase shifts from UChPT~\cite{Liu:2012zya}, which
contains the $D_0^*(2100)$, and the BW for the $D_0^*(2300)$.
The fit results are shown in Fig.~\ref{fig:KT}, where the blue and green bands correspond to the best fits
from UChPT and the BW, respectively.
While UChPT describes the data very well with $\chi^2/\text{d.o.f.}=1.2$,
the BW fails to reproduce the data with $\chi^2/\text{d.o.f.}=2.0$. 
The difference of these two values is significant from the statistical point of view: the corresponding $p$-values are 0.1 for the UChPT fit and $3\times10^{-5}$ for the BW fit, respectively. Thus, the former can be accepted as a good description of the data, while the latter is highly disfavored~\cite{Press:1992zz}.
The error bands
correspond to the $1\sigma$ uncertainties propagated from the input phases. 
The borders of the band of $\langle P_{13}\rangle$ for the BW are plotted in dotted and dashed curves to
make it evident that the data for $\langle P_{0}\rangle$ and $\langle P_{13}\rangle$ cannot be described by the 
BW phase.
With the fitted parameters $g_0$ and $g_1$, we obtain the $S$-wave phase of the decay amplitude for
$B^-\to D^+\pi^-\pi^-$ shown in Fig.~\ref{fig:phaseIV}, where the results corresponding to UChPT and the BW
are plotted as blue and green bands, respectively. As expected, UChPT describes the $S$-wave phase
extracted using Eq.~\eqref{eq:cos} and that obtained in Ref.~\cite{Aaij:2016fma} well up to 2.4\GeV. For the BW
one, although the error band is broad, either the low-energy or the high-energy region cannot be described.

{\it Conclusion.}---The existence of the $D_0^*(2300)$ as given in the RPP is the starting point of many
theoretical analyses (see, e.g., Refs.~\cite{Mehen:2005hc,Colangelo:2012xi,Alhakami:2016zqx,Cheng:2017oqh}). 
The results obtained in this Letter show that the $D_0^*(2300)$, whose resonance parameters were obtained
using the BW parametrization from the Belle~\cite{Abe:2003zm} and BaBar~\cite{Aubert:2009wg} analyses, is
in conflict with the much more precise LHCb data for $B^-\to D^+\pi^-\pi^-$, which, however, can be well
reproduced by the UChPT amplitude containing the $D_0^*(2100)$.

We expect that the $D_1(2430)$ as given in the RPP~\cite{Zyla:2020zbs} will also be in conflict with
future high-quality data of $B^-\to D^{*+}\pi^-\pi^-$ from LHCb~\cite{Aaij:2019sqk} and Belle-II, and that the lightest $D_1$ meson
is the $D_1(2250)$ predicted by UChPT~\cite{Albaladejo:2016lbb,Du:2017zvv}.

The $D_0^*$ is analogous to the more famous $f_0(500)$ and $K_0^*(700)$, whose masses have
been significantly shifted from earlier versions of the RPP due to improved data
and improved theoretical analyses---for recent discussions see
Refs.~\cite{Caprini:2005zr,DescotesGenon:2006uk,Pelaez:2015qba,Pelaez:2020uiw,Pelaez:2020gnd} and the review on scalar mesons in 
the RPP~\cite{Zyla:2020zbs}.
We expect a similar change in all systems emerging from the scattering of a pion 
off an
isospin-nonsinglet hadron. The lightest resonance in that case should not be extracted from data using
the usual BW form---a parametrization accounting for chiral symmetry and coupled channels is mandatory.
The $D\pi$ $S$-wave phase extracted model-independently here provides valuable information for further understanding matter-field--Goldstone-boson scattering and the structure of positive-parity heavy hadrons.

{\it Note added.}---Recently,  
a lattice calculation also concluded that the $D_0^*$ mass should be lower than the RPP value~\cite{Gayer:2021xzv}. The authors found a mass of $(2196\pm64)\MeV$ with a pion mass of $239\MeV$, only $(77\pm64)\MeV$ above the $D\pi$ threshold.
Thus, our conclusion receives a strong support from lattice QCD calculations. 

\smallskip

\begin{acknowledgments}
This work is supported in part by the National Natural Science Foundation of China (NSFC) under Grants No.~11835015,
No.~12047503, and No.~11961141012, by the NSFC and the Deutsche Forschungsgemeinschaft (DFG, German Research
Foundation) through the funds provided to the Sino-German Collaborative
Research Center ``Symmetries and the Emergence of Structure in QCD''
(NSFC Grant No.~12070131001, DFG Project-ID 196253076 -- TRR110), by the Chinese Academy of Sciences (CAS) under Grants No.~XDB34030000
and No.~QYZDB-SSW-SYS013, and by the CAS Center for Excellence in Particle Physics (CCEPP). The work of
U.G.M. was also supported by the Chinese Academy of Sciences (CAS) President's International Fellowship
Initiative (PIFI) (Grant No.~2018DM0034), by VolkswagenStiftung (Grant No.~93562), and by the EU (Strong2020).
\end{acknowledgments}

\bibliographystyle{apsrev}
\bibliography{dpi}

\begin{onecolumngrid}
\appendix
\newpage
\begin{center}
{ \it \large Supplemental Material}\\
\vspace{0.05in}
\end{center}

\section{Technical details}

The inhomogeneity for the $S$-wave $s$-channel amplitude is
\be 
\hat{\mathcal{F}}_0^{1/2}(s) = -\frac{1}{4\sqrt{2}}\int_{-1}^1 dz_s(t-s)\mathf_1^1(u)  
 =  -\frac{s}{4\sqrt{2}\kappa(s)}\int_{u_-(s)}^{u_+(s)}du(\Sigma_0-u-2s)\mathf_1^1(u),
\ee
where $\Sigma_0=M_B^2+M_D^2+2M_\pi^2$, and the integral end points and path need to be understood with the prescription $M_B^2\to M_B^2+i\epsilon$.
 The inhomogeneities give the contributions from the crossed 
 channels, i.e., the left-hand cut contributions. 
 The integration end points $u_\pm(s)$ contain the only nontrivial $M_{B}$-dependent pieces. The continuation following the $M_{B}^2 \to M_{B}^2 +i\epsilon$ prescription gives
\bea
u_+(s) \al = \al \left\{ \begin{array}{ll}
\frac{\Sigma_0s-s^2-\Delta+|\kappa_s(s)|+i\epsilon}{2s}, & s\in [(M_D+M_\pi)^2,(M_B-M_\pi)^2), \\
\frac{\Sigma_0s-s^2-\Delta+i|\kappa_s(s)|}{2s}, & s\in [(M_B-M_\pi)^2,(M_B+M_\pi)^2), \\
\frac{\Sigma_0s-s^2-\Delta-|\kappa_s(s)|}{2s},& s\in [(M_B+M_\pi)^2,\infty ),
\end{array}
\right.\nonumber \\
u_-(s) \al = \al \left\{ \begin{array}{ll}
\frac{\Sigma_0s-s^2-\Delta-|\kappa_s(s)|+i\epsilon}{2s}, & s\in [(M_D+M_\pi)^2,(M_{B}^2+M_D^2-2M_\pi^2)/2), \\
\frac{\Sigma_0s-s^2-\Delta-|\kappa_s(s)|-i\epsilon}{2s}, & s\in \Big[(M_{B}^2+M_D^2-2M_\pi^2)/2,(M_B-M_\pi)^2\Big),\\
\frac{\Sigma_0s-s^2-\Delta-i|\kappa_s(s)|}{2s}, & s\in [(M_B-M_\pi)^2,(M_B+M_\pi)^2), \\
\frac{\Sigma_0s-s^2-\Delta+|\kappa_s(s)|}{2s},& s\in [(M_B+M_\pi)^2,\infty ).
\end{array}
\right. \nonumber
\eea

The amplitude $\mathf_1^1(u)$ is parameterized by a Breit-Wigner (BW) amplitude of the $\rho^-$ meson, 
since, as explained in the main text,
the process $B^-\to D^0\pi^0\pi^-$ is assumed to proceed through the intermediate 
decay $B^-\to D^0\rho^-$, 
\bea\label{eq:F11u}
\mathf_1^1(u)  = c_S R_1(u)\times X_1( |\vec{p}_u| r_\text{BW}) \times X_1( |\vec{q}_u| r_\text{BW}),
\eea
where the $\vec{p}_u$ and $\vec{q}_u$ are the momenta of the $D^0$ and $\pi^+$ ($\pi^0$) in the 
$\pi^0\pi^-$ rest frame, respectively. The $\rho$ resonance lineshape is modelled by a relativistic BW 
function $R_1(u)$ given by
\bea
R_1(u) = \frac{1}{M_\rho^2-u-i M_\rho \Gamma({u})},
\eea
with a mass-dependent decay width defined as 
\bea
\Gamma({u}) = \Gamma_\rho \Big( \frac{|\vec{q}_u|}{|\vec{q}_u^0|}\Big)^3 \frac{M_\rho}{\sqrt{u}} X_1^2(|\vec{q}_u| r_\text{BW}),
\eea
where $q_u^0$ is the value of $q_u$ when $\sqrt{u}=M_\rho$, and $\Gamma_\rho$ is the full width. 
The Blatt-Weisskopf barrier factors $X_L(x)$ are given by
\begin{equation}
X_0(x)  =  1, \qquad
X_1(x)  =  \sqrt{ \frac{x_0^2+1}{x^2+1} }, \qquad
X_2(x)  =  \sqrt{ \frac{ x_0^4+ 3x_0^2+9}{ x^4+3x^2+9}} ,
\end{equation}
where $x = |\vec{q}|r_\text{BW}$ or $|\vec{p}|r_\text{BW}$, and $x_0$ is the value of $x$ evaluated 
at $\sqrt{u}=M_\rho$. The barrier radius, $r_\text{BW}$, is taken to be $4.0\GeV^{-1}$ to be consistent with 
Ref.~\cite{Aaij:2016fma}. 

For the $P$-($D$-)wave amplitudes $\mathf_{1,2}^{1/2}(s)$, we use the same BW amplitudes as those 
in the LHCb analysis~\cite{Aaij:2016fma}, i.e.,
\bea\label{eq:bwJ} 
\mathf_J^{1/2}(s)\al = \al c_J R_J(s)\times X_J( |\vec{p}|r_\text{BW}) \times X_J(|\vec{q}|r_\text{BW})(-s)^J,
\eea
where $\vec{p}$ and $\vec{q}$ are the momenta of the $\pi^-$($p_2$) and the $\pi$($p_1$), 
calculated in the $D\pi(p_1)$ rest frame. $R_J(s)$ is given by 
\be
R_J(s)  =  \frac{1}{m_J^2-s-im_J\Gamma_J({s})}, \qquad
\Gamma_J({s})  =  \Gamma_J^0 \Big( \frac{|{\vec{q}}|}{|\vec{q}_0|}\Big)^{2J+1} \Big(\frac{m_J}{\sqrt{s}}\Big)X_J^2(|\vec{q}| r_\text{BW}), 
\ee
where $\vec{q}_0$ is the value of $\vec{q}$ evaluated at the resonance mass $\sqrt{s}=m_J$ with $m_J$ the mass of the $P$- or $D$-wave resonance and $\Gamma_J^0$ the corresponding width. The $P$-wave resonance $D^*$ is outside the kinematically allowed region. Therefore, its mass
$m_1$ is modified to an effective mass $m_1^\text{eff}$ as in the LHCb analysis~\cite{Aaij:2016fma},
\bea
m_1^\text{eff} = m^\text{min}+(m^\text{max}-m^\text{min})\Bigg( 1+ \text{tanh}\frac{m_1-\frac{m^\text{min}+m^\text{max}}{2}}{m^\text{max}-m^\text{min}}\Bigg),
\eea
where $m^\text{max}$ and $m^\text{min}$ represent the upper and lower limits of the kinematically allowed range.

\section{Parameters}

The masses and widths of the $P$- and $D$-wave resonances used in the calculation are listed in Table~\ref{tab:masswidth} and are the same as those used in the LHCb analysis~\cite{Aaij:2016fma}. The UChPT amplitudes for the $D\pi$ interaction and corresponding low energy constants (LECs) used in this work 
are taken from Ref.~\cite{Liu:2012zya}. For the BW parameterization of the $S$-wave $D\pi$ amplitude, the amplitude of Eq.~\eqref{eq:bwJ} with $J=0$ is employed. 
\begin{table}[htb]
\centering
\caption{\small Masses and widths of the vector and tensor charmed mesons as used in Ref.~\cite{Aaij:2016fma}.}
\label{tab:masswidth}
\renewcommand{\arraystretch}{1.3}
\begin{tabular}{lccc}
\hline
\hline 
Contribution & Spin & Mass (MeV) & Width (MeV) \\ 
\hline 
$D^\ast(2007)^0$ & 1 &   $2006.98$ & 2.1\\ 
$D_2^\ast(2460)^0$ & 2 &   $2463.7 $ & 47.0\\ 
\hline
\hline 
\end{tabular}
\end{table}

The parameters determined in the best fits to data for the Khuri-Treiman formalism, e.g., Eq.~\eqref{eq:Fspcut}, with input from UChPT and the BW amplitudes, respectively, are collected in Table~\ref{tab:paras}, where only the statistical uncertainties are given. Each partial wave has an overall normalization factor, e.g., $c_1$ and $c_2$ for $P$- 
and $D$-waves, respectively. Since we fit to the unnormalized angular moments, the units of the fitted parameters 
contain some arbitrary overall normalization factors. 

\begin{table}[htb]
\centering
\caption{\small Parameters determined in the best fit to data for the Khuri-Treiman formalism with input from UChPT and the BW amplitudes, respectively, with statistical uncertainties only. Here, the $g_i=a_ie^{i\delta_i}$ refer to the subtraction constants in Eq.~\eqref{eq:Fspcut}, with $a_i$ real numbers.}
\label{tab:paras}
\renewcommand{\arraystretch}{1.3}
\begin{tabular}{lcc}
\hline
\hline 
Parameters & UChPT & BW \\ 
\hline 
$10^3\cdot a_0$~ & $0.57\pm0.14$ & $0.59\pm 0.25$ \\
$\delta_0$ & $-0.38\pm0.08$ & $-2.57 \pm 0.05$ \\
$10^3\cdot a_1$~ & $ 1.15\pm 0.75$ & $0.95 \pm 0.38$ \\
$\delta_1$& $1.56\pm 0.40$ & $-0.15\pm 0.05$ \\
$10^{3} \cdot c_S $~& $1.36\pm0.46$ & $0.00\pm 0.00$\\
$10^{4} \cdot c_1 $~ & $1.65\pm0.06$  & $1.56\pm 0.05$ \\
$10^{4} \cdot c_2 $~[GeV$^{-2}$] & $1.47\pm0.07$ & $1.64\pm0.06$ \\
\hline
\hline 
\end{tabular}
\end{table}

\section{Describing the data in a two-body isobar model}

To make it explicit that the requirement of chiral symmetry has a sizeable impact on the determination of resonance parameters using the BW parameterization in the isobar model as argued in Ref.~\cite{Du:2019oki}, here we show two different fits:
\begin{itemize}
\item model~I: the $S$-wave $D\pi$ is parameterized as a relativistic BW amplitude, Eq.~\eqref{eq:bwJ};
\item model~II: the $S$-wave $D\pi$ is parameterized as a BW amplitude times an $E_\pi$ factor, as required by the 
chiral symmetry of QCD:
\bea
\mathf_0^{\prime}(s) = c_0 R_0(s)\times 2E_\pi/M_D,
\eea
with $E_\pi$ the energy of the pion in the $D\pi$ rest frame.
\end{itemize}
Note that in the isobar model, the crossed-channel effects are not explicitly considered and the coefficients 
$c_J$ are complex to describe the relative contribution of each intermediate process. Here, we do not use the resonance parameters of $D_0^*$ in the Review of Particle Physics (RPP), but determine the BW
mass and width by fitting to the high quality LHCb data in Ref.~\cite{Aaij:2016fma}. The best fits to the angular moments 
are shown in Fig.~\ref{fig:bw}, where the orange (with $\chi^2/\text{d.o.f.} = 1.8$) and the blue curves ($\chi^2/\text{d.o.f.} = 1.1$) correspond to  model~I and model~II, respectively. For model~I, the fit results in the $D_0^*$ BW mass and width parameters 
of $(2319\pm 9)\MeV$  and $(420\pm 25)\MeV$, respectively. The fitted mass is consistent with 
the RPP average mass of $(2300\pm 19)$ MeV. However, the width is much larger than the RPP average value, 
i.e., $(274\pm 40)\MeV$. 
Note that in Ref.~\cite{Aaij:2016fma}, a quasi-model-independent approach is used to describe the $D^+\pi^-$ $S$-wave. The $D^+\pi^-$ $S$-wave is extracted using cubic splines to describe the magnitude and phase variation. Accordingly, in this reference no pole parameters for the
$D_0^*$ are given.
When the $E_\pi$ factor is considered, i.e., in model~II, the fit quality is significantly improved with a much smaller 
$\chi^2$. The fitted parameters are given in Table~\ref{tab:isobar} with only statistical uncertainties. In addition, the fit results in a BW mass of $(2206\pm 4)\MeV$, which is much lower than that given in the RPP, while the width is $(341\pm 17)\MeV$.
This clearly shows that the BW parameters for a broad resonance can be affected significantly by the chiral symmetry constraint, as argued in Ref.~\cite{Du:2019oki}. Nevertheless, the modification of the BW in model~II only applies in a small energy region before the coupled-channel 
effects become important, and thus is neither practical nor systematic. 
The resonance parameters would get further modified after the $D\eta$ and $D_s\bar K$ coupled channels are taken into account.
The UChPT formalism used in the main text provides a theoretical framework satisfying both chiral symmetry and coupled-channel unitarity.

\begin{figure*}[tb]
\includegraphics[width=0.32\textwidth]{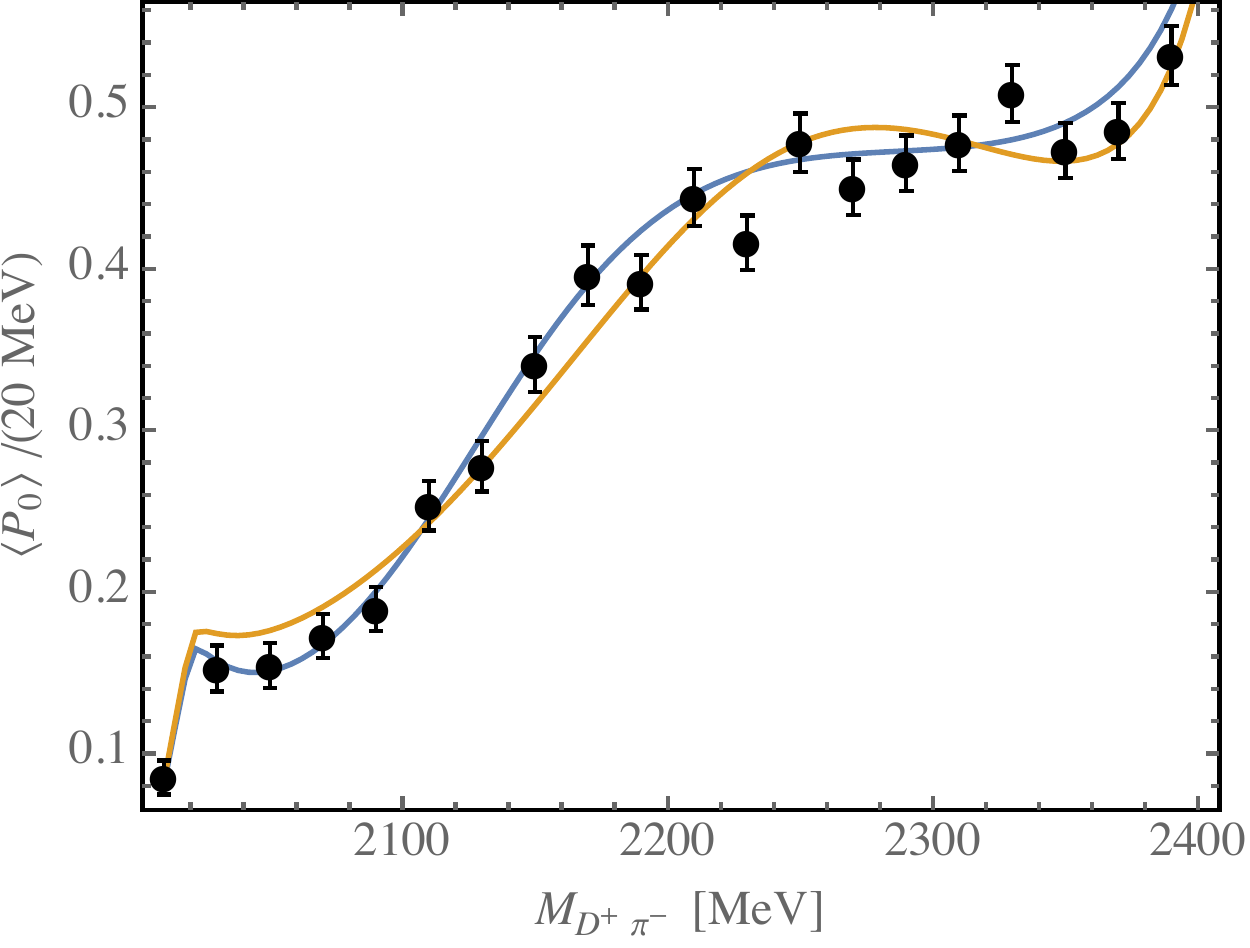}
\includegraphics[width=0.32\textwidth]{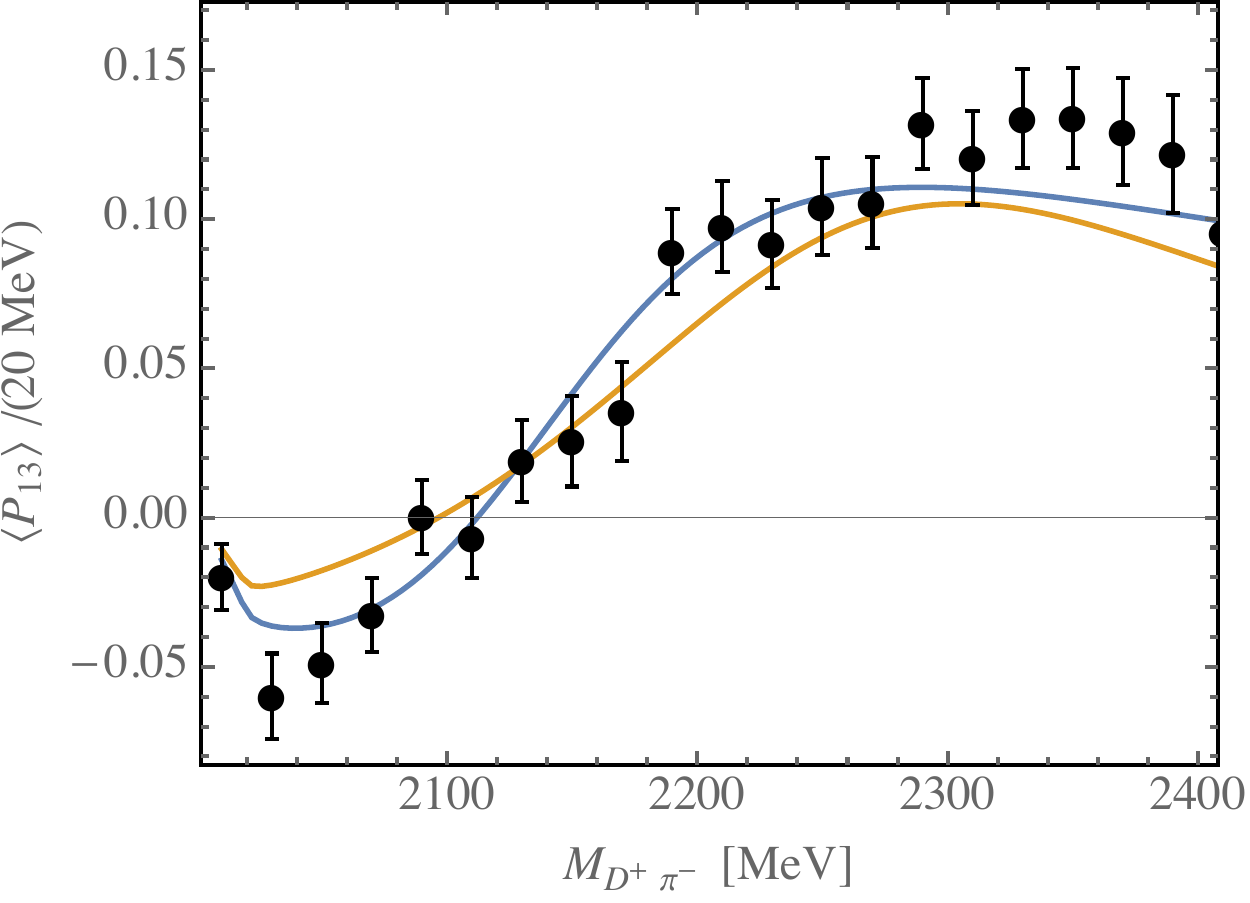}
\includegraphics[width=0.32\textwidth]{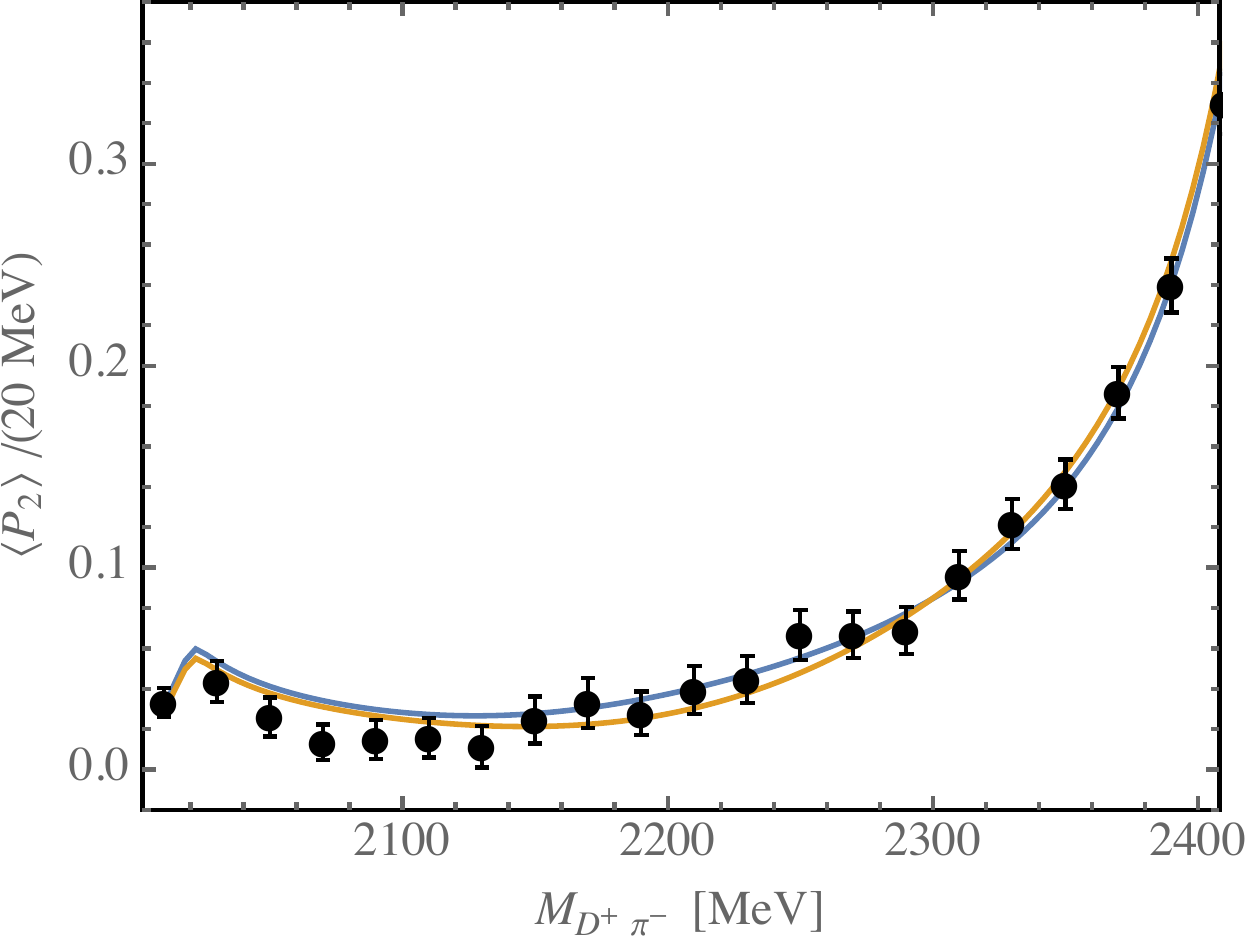}
\caption{The best fit results of  model~I (orange) and model~II (blue) in the isobar model, with $\chi^2/{\rm d.o.f.} = 1.8$ and 1.1, respectively.}\label{fig:bw}
\end{figure*}

\begin{table}[htb]
\centering
\caption{\small Parameters determined by the fit to data in the two-body isobar model for model~I and II with statistical uncertainties only, where $c_i = |c_i|e^{i\phi_i}$ with $\phi_i$ real numbers and the phase of $D$-wave is fixed to zero, i.e., $\phi_2=0$. 
}
\label{tab:isobar}
\renewcommand{\arraystretch}{1.3}
\begin{tabular}{lcc}
\hline
\hline 
Parameters & model~I & mdoel~II \\ 
\hline 
$10^{3}\cdot |c_0| $ ~[GeV$^{2}$] & $3.39\pm0.18$ & $7.28\pm 0.30$ \\
$\phi_0$ & $1.08\pm0.17$ & $0.98 \pm 0.21$ \\
$10^4\cdot |c_1| $~~  & $ 1.60\pm 0.05$ & $1.67 \pm 0.05$ \\
$\phi_1$& $-0.08\pm 0.16$ & $0.30\pm 0.21$ \\
$10^{4} \cdot |c_2| $ ~[GeV$^{-2}$] & $1.71\pm 0.11 $ & $1.43\pm 0.07$\\
$m_\text{BW} $ ~~~~~[GeV]& $2.319\pm 0.009$  & $2.206\pm 0.004$ \\
$\Gamma_\text{BW}^0$ ~~~~~~[GeV] & $0.420\pm 0.025$ & $0.341\pm 0.017$ \\
\hline
\hline 
\end{tabular}
\end{table}

\end{onecolumngrid}

\end{document}